\documentclass[journal]{IEEEtran}

\usepackage{cite}

\usepackage[pdftex]{graphicx}
\graphicspath{{./}}
\DeclareGraphicsExtensions{.pdf,.jpeg,.png,.jpg,.eps}

\usepackage[cmex10]{amsmath}
\usepackage{array,bm}
\usepackage{multirow}
\usepackage{url}
\usepackage{booktabs}
\usepackage{float}
\usepackage{mathtools}
\usepackage{amsfonts}
\usepackage{diagbox}
\usepackage{amsmath}
\usepackage{amssymb}
\usepackage{mathrsfs}
\usepackage{threeparttable}
\usepackage{subfigure}
\usepackage{tikz}

\newcommand*\circled[1]{\tikz[baseline=(char.base)]{
		\node[shape=circle,draw,inner sep=0.8pt] (char) {#1};}}

\begin{document}

\title{Data-Driven Transient Stability Boundary Generation for Online Security Monitoring}

\author{
	    Rong~Yan,~\IEEEmembership{Graduate Student Member,~IEEE,}
	    Guangchao~Geng,~\IEEEmembership{Senior Member,~IEEE,}
        and~Quanyuan~Jiang,~\IEEEmembership{Senior Member,~IEEE} 

\thanks{All the authors are with the College of Electrical Engineering, Zhejiang University, Hangzhou, Zhejiang 310027, China (email: \{yanrong052, ggc, jqy\}@zju.edu.cn). \textit{(Corresponding author: G. Geng.)}}}

\markboth{SUBMITTED TO JOURNAL FOR POSSIBLE PUBLICATION. COPYRIGHT MAY BE TRANSFERRED WITHOUT NOTICE}
{Yan \MakeLowercase{\textit{et al.}}: Data-Driven Transient Stability Boundary Generation for Online Security Monitoring}

\maketitle

\begin{abstract}
Transient stability boundary (TSB) is an important tool in power system online security monitoring, but practically it suffers from high computational burden using state-of-the-art methods, such as time-domain simulation (TDS), with numerous scenarios taken into account (e.g., operating points (OPs) and N-1 contingencies). The purpose of this work is to establish a data-driven framework to generate sufficient critical samples close to the boundary within a limited time, covering all critical scenarios in current OP. Therefore, accurate TSB can be periodically refreshed by tracking current OP in time. The idea is to develop a search strategy to obtain more data samples near the stability boundary, while traverse the rest part with fewer samples. To achieve this goal, a specially designed transient index sensitivity based search strategy and critical scenarios selection mechanism are proposed, in order to find out the most representative scenarios and periodically update TSB for online monitoring. Two case studies validate effectiveness of the proposed method. 

\end{abstract}


\IEEEpeerreviewmaketitle

\section{Introduction}
\IEEEPARstart{T}{ransient} stability is an important issue in power system planning, operation and control, as it is one of the most primary considerations in system security. In these applications, it is a regular routine to establish a description of transient stability boundary (TSB), no matter accurate or approximate, by performing time-domain simulation (TDS) for numerous operating points (OPs) and possible contingencies. Computational burden of TDS is especially heavy, if not unacceptable, after data-driven techniques have been employed to find out TSB. Thus, the key challenge is to improve the efficiency of obtaining sufficient TDS samples near TSB with representative scenarios for accurate boundary generation.

In general, most existing works are based on model-based approaches, like TDS \cite{Adibi1974Solution}, transient energy function \cite{Chang1995Direct} and extended equal area criteria \cite{Xue1998Fast}, but with certain drawbacks of computational burden, model adaptability, or reliability, respectively. The advent of data-driven technique provides an alternative way to relieve such difficulties. A huge amount of transient simulation data are, however, required to construct a reliable stability boundary, as there is a lack of accumulated historical data with disturbances during daily operation, in terms of both variety and severity. Monte Carlo Sampling with TDS is applied in most existing works, changing all loads randomly within a predetermined range uniformly \cite{He2013Robust}, \cite{WangBo_whu}, \cite{lstm1}, \cite{cnn7} or independently \cite{yan2019}. However, generators outputs, in those works, are then scheduled by a certain rule (e.g., optimal power flow), to emulate practical system operations. Considering that the actual OP may not meet such rules, a good point set method \cite{xuyan_sampling} is proposed to improve the solution.

After realizing the low efficiency of the methods aforementioned, re-sampling based techniques are introduced in this area \cite{resampling1}, \cite{resampling2}, \cite{resampling3} to enrich scenarios based on the existing generated data. Among them, binary search algorithm is employed in \cite{resampling1} to enhance the data set iteratively. Besides, importance sampling technique is also applied in an iterative framework \cite{resampling2}, \cite{resampling3} to identify the more important (or ``informative'') region prior to TDS. Such efforts have been made to improve the quality of training data set for TSB.

Therefore, the goal of this work is to obtain sufficient TDS data near TSB with most critical scenarios and generate boundary by tracking current OP efficiently, which is necessary in dynamic security online monitoring and control. This is rarely addressed in the existing works, but exactly the key technology of data-driven based online security monitoring. Inspired by the adjoint sensitivity analysis (ASA) from dynamic optimization methods like \cite{lizhihao}, we propose to explore critical OPs close to the TSB, following the guidance of first-order derivative information of the specially designed transient index, so as to improve the information entropy of data samples. In addition, high-dimension variables clustering technique is also employed to find out the data relevance among the combinations of OPs and contingencies, in order to reduce the search space. As more data samples with representative OPs and contingencies accumulates, TSB can be constructed more accurately and efficiently.

According to this motivating idea, a data-driven transient stability boundary generation (DTSBG) framework for online security monitoring is proposed in this paper. As a continuation of the authors’ previous work of early terminating TDS for transient stability batch assessment \cite{yan2019}, this paper is highlighted with the following contributions:

\begin{enumerate}
	\item A critical data sampling framework with adjoint sensitivity based index, enforcing sampled data close to TSB;
	\item A re-sampling method to fill more data in gap area of TSB generation, providing more samples across TSB;
	\item A critical scenario selection strategy to identify the relevance of scenario set and relief computational burden.
\end{enumerate}

\section{Transient Stability Boundary Generation}

\subsection{Transient Stability Boundary Formulation}
Stability boundary can be formulated as a set of constraints shown as follows:
\vspace{-3pt}
\begin{equation}
	\label{eq:DAE0}
	\left\{
	\begin{aligned}
		0&=\bm{P}(u)\\
		\underline{\bm{J}} &\leq \bm{J}(u) \leq \overline{\bm{J}}\\
		0&=\bm{Z}(u,x_0)\\
		\underline{\bm{\Phi}}&\leq\bm{\Phi}^k(u,\bm{x}^k(t),\bm{y}^k(t))\leq\overline{\bm{\Phi}}
	\end{aligned}
	\right.
	\vspace{-3pt}
\end{equation}
where $\bm{P}$ are the power flow equations, $\bm{J}$ are the static operation constraints, $\bm{Z}$ are the initial condition equations, and $\bm{\Phi}^k$ are the transient constraints for the $k$-th contingency. $\overline{\bullet}$ and $\underline{\bullet}$ are the upper and lower bound of the variable. While, as for variables in Eq. (\ref{eq:DAE0}), $u$ are static variables of the given operation point, including active and reactive power of generators, and amplitude and phase angle of voltages. $x_0$ are the initial values of state variables. $\bm{x}^k$ and $\bm{y}^k$ are the time-variant state variables and operating variables respectively that describe dynamic response of the system after the $k$-th contingency by solving the differential equations $\bm{F}^k$ and algebraic equations $\bm{G}^k$ shown in the following equations:
\vspace{-3pt}
\begin{equation}
	\label{eq:TDS}
	\left\{
	\begin{aligned}
		\dot{\bm{x}}^k(t)& = \bm{F}^k(u,\bm{x}^k(t),\bm{y}^k(t)) \\
		\bm{0~} & = \bm{G}^k(u,\bm{x}^k(t),\bm{y}^k(t))\\
		\bm{x}^k(0)&=x_0
	\end{aligned}
	\right.
	\vspace{-3pt}
\end{equation}
As a matter of fact, only the last equation in Eq.(\ref{eq:DAE0}) is time-consuming while formulating the boundary because of the simulation part in Eq.(2). Therefore, in this paper, we only focus on the transient constraint or boundary of this type.

As for the transient stability inequality constraint, there are several ways to formulate. In this paper, we choose the absolute deviation of rotor angle with respect to the center of inertia (COI). It is shown in Eq.(\ref{eq:rotor_constraint}),  
\vspace{-3pt}
\begin{equation}
	\label{eq:rotor_constraint}
	-\delta_{max} \leq \delta_i^k(t|u)-\delta_{COI}^k(t|u) \leq \delta_{max}, ~~~t\in[0,T]
	\vspace{-3pt}
\end{equation}
where $i=1,2,...,N_G$, $N_G$ stands for the number of generators in the given power system, $T$ is the length of simulation time window while solving the DAEs defined in Eq. (\ref{eq:TDS}), $\delta_i^k(t)$ denotes the absolute rotor angle of the $i$-th generator under the $k$-th contingency at time $t$, and $\delta_{COI}^k(t)$ indicates COI which is defined as:
\vspace{-3pt}
\begin{equation}
	\label{eq:coi}
	\delta_{COI}^k(t|u)=\frac{1}{\sum_{i=1}^{N_G} T_{Ji}}\sum_{i=1}^{N_G} T_{Ji}\delta_i^k(t|u)
	\vspace{-3pt}
\end{equation}
where $T_{Ji}$ denotes inertia time constant of the $i$-th generator.

\vspace{-8pt}
\subsection{Transient Stability Index and its Sensitivity}
Since it is difficult to handle Eq. (\ref{eq:rotor_constraint}) directly, the constraint transformation technique is applied here to define a equivalent transient stability index as follows:
\vspace{-3pt}
\begin{equation}
    \small
	\label{eq:Transient_stability_index1}
	\bm{\Phi}^k(u)  =\int_0^T\max\{ 0,\theta^k[\delta^k(t|u)]\}dt
	                    =\int_0^T\widetilde{\theta}^k[\delta^k(t|u)]dt
	\vspace{-3pt}
\end{equation}
where $\theta^k[\delta^k(t|u)]$ is defined as:
\vspace{-3pt}
\begin{equation}
	\label{eq:Transient_stability_index2}
	\theta^k[\delta^k(t|u)]=\varLambda \cdot (\delta_{max}^2-max\{(\delta^k_i(t|u)-\delta^k_{COI}(t|u))^2\})
	\vspace{-3pt}
\end{equation}
In Eq.(\ref{eq:Transient_stability_index2}), $\varLambda$ indicates a constant, in order to distinguish the stable scenario from unstable one, and it is given by
\vspace{-3pt}
\begin{equation}
	\label{eq:Transient_stability_index3}
	\varLambda=
	\left\{
		\begin{array}{ccc}
			1     &  & \text{Stable OP}\\
			-1    &  & \text{Unstable OP}\\
		\end{array}
	\right.
	\vspace{-3pt}
\end{equation}
Therefore, $\bm{\Phi}^k(u)$ can be seen as an index to measure the dynamic performance of a given OP.

To guide the generated OPs near the stability boundary, it is of significance to obtain the first-order derivative information of the stability index $\bm{\Phi}$. To do so, forward and adjoint sensitivity analysis are two major approaches. The latter one is preferred due to its lower computational burden. The first-order derivative information can be calculated as Eq. (\ref{eq:gradient}).
\vspace{-3pt}
\begin{equation}
	\label{eq:gradient}
	\begin{split}
		\triangledown_u \bm{\Phi}^k(u)&=\frac{\partial{\bm{\Phi}^k(u)}}{\partial u}\\
		                  &=\int_0^T \frac{\partial \bm{H}^k[u,\bm{x}^k(t),\bm{y}^k(t), \bm{\lambda}^k(t),\bm{\beta}^k(t)]}{\partial u}dt\\                 
	\end{split}
	\vspace{-3pt}
\end{equation}
where $\bm{H}^k[u,\bm{x}^k(t),\bm{y}^k(t), \bm{\lambda}^k(t),\bm{\beta}^k(t)]$ is the Hamilton function corresponding to the transient constraint for the $k$-th contingency, and can be defined as follows:
\vspace{-3pt}
\begin{equation}
	\label{eq:hamilton}
	\begin{split}
		\bm{H}^k(u,\bm{x}^k,\bm{y}^k, \bm{\lambda}^k,\bm{\beta}^k)&=\widetilde{\theta}^k+(\bm{\lambda}^k)^T \bm{F}^k (u,\bm{x}^k,\bm{y}^k)\\
		&+(\bm{\beta}^k)^T \bm{G}^k (u,\bm{x}^k,\bm{y}^k)
	\end{split}
	\vspace{-3pt}
\end{equation}
where $\bm{\lambda}^k$ and $\bm{\beta}^k$ are the time solution to co-state equation:
\vspace{-3pt}
\begin{equation}
	\label{eq:co_state}
	\left\{
	\begin{aligned}
		\dot{\bm{\lambda}}^k(t)& = -\frac{\partial \bm{H}^k[u,\bm{x}^k(t),\bm{y}^k(t), \bm{\lambda}^k(t),\bm{\beta}^k(t)]}{\partial \bm{x}^k}\\
		\bm{0~} & = \frac{\partial \bm{H}^k[u,\bm{x}^k(t),\bm{y}^k(t), \bm{\lambda}^k(t),\bm{\beta}^k(t)]}{\partial \bm{y}^k}\\
		\bm{0~} &=\bm{\lambda}^k(T)
	\end{aligned}
	\right.
	\vspace{-3pt}
\end{equation}
The detailed proof of Eq.(\ref{eq:gradient})-(\ref{eq:co_state}) can be referred to \cite{lizhihao}.

\vspace{-8pt}
\subsection{Critical Data Samples Sampling Strategy}
\begin{figure}
	\vspace{-18pt}
	\centering
	\includegraphics[width=3.5in]{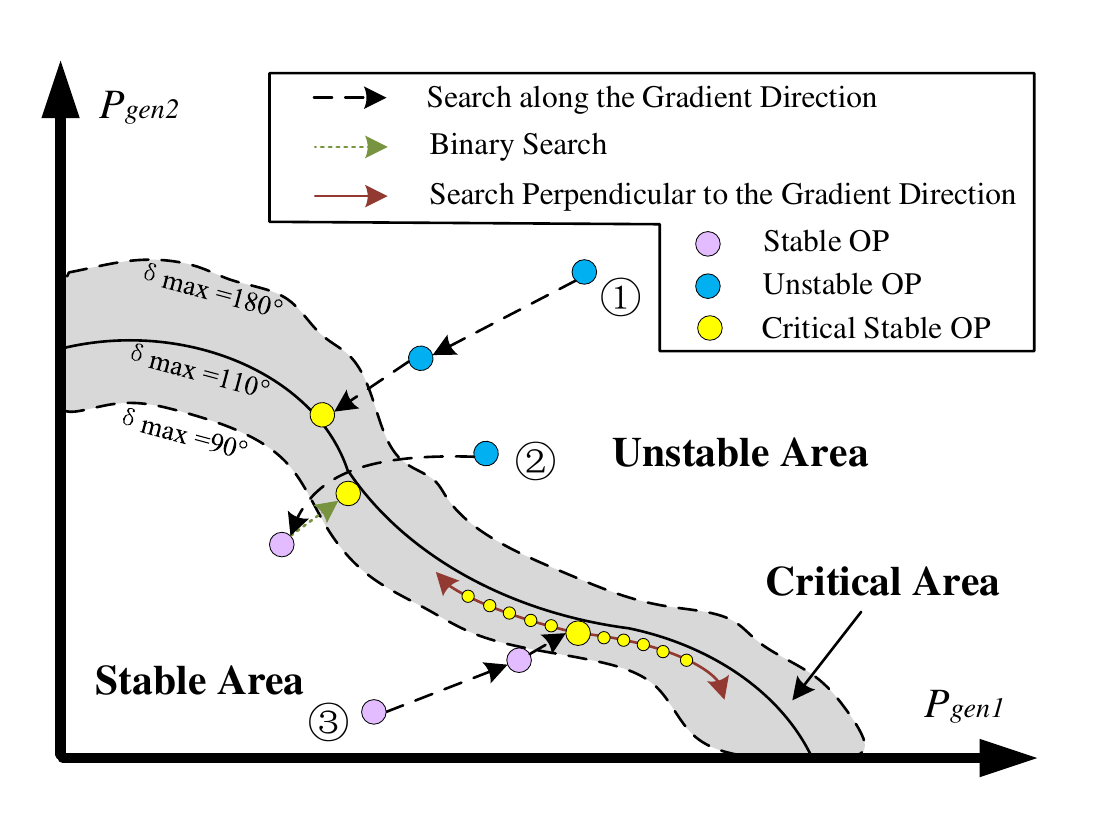}
	\vspace{-25pt}
	\caption{Diagram of critical data sampling strategy.}
	\label{fig:search}
	\vspace{-12pt}
\end{figure}
In order to obtain more data samples close to the boundary, three main indices are required: one measures the distance between current sampled OP and boundary, the other two are the direction and step size guiding the current OP moving close to the boundary. In terms of distance measurement, we choose $\bm{\Phi}^k(u)$ reflecting how stable/unstable the system is. In terms of direction, the first-order derivative information $\triangledown_u \bm{\Phi}^k(u)$ is employed. While, for step size, we design a variable step algorithm, in order to move the sampling OPs near the stability boundary as fast as possible. In that case, more critical OPs near the TSB can be sampled and simulated. Considering that larger step size can be set when the current OP is far from the boundary, while the opposite is smaller. Therefore, the step size $\zeta_m$ is can be calculated as:
\vspace{-3pt}
\begin{equation}
	\label{eq:step_size}
	\zeta_m=\nu_{i,j} \cdot u_{max}
	\vspace{-3pt}
\end{equation}
where $u_{max}$ indicates a vector consisting with the maximum value of the controllable variables in the OPs, while $\nu_{i,j}$ is the coefficient with respect to the stability index $\bm{\Phi}^k(u)$. Thus, the next sampling OPs can be calculated as follows:
\vspace{-3pt}
\begin{equation}
	\label{eq:step}
	u_{m+1}=u_{m}-\zeta_m\cdot\bigtriangledown_u\bm{\Phi}^k(u)
	\vspace{-3pt}
\end{equation}

After repeating this step until the new OP is found to be a critical one (close to the boundary), just as the route \circled{1} shown in Fig \ref{fig:search}. However, some search processes cross critical area as the route \circled{2} considering highly nonlinear character of transient stability problem, a small trick like binary search is employed to handle such cases. Additionally, the new transient samples can also be positioned perpendicular to the gradient direction with small step size based on the existing samples near the boundary, as the route \circled{3}, to further enrich transient dataset close the boundary. Throughout the data collection process of these three routes, data samples with high information entropy are obtained for data-driven security assessment.

\vspace{-8pt}
\subsection{Boundary Generation and Gap Area Re-sampling}
After obtaining sufficient data samples using the strategy in previous subsection, TSB can be generated using data-driven algorithms, like decision trees, support vector machine, k-nearest neighbors algorithm, convolutional neural network, and etc. No matter what exact algorithm we choose, the essence of the boundary can be summarized as a complicated function with respect to the OP $u$ and contingency $k$ given as follows:
\vspace{-3pt}
\begin{equation}
	\label{eq:boundary}
	\varGamma(u,k)=0 
	\vspace{-3pt}
\end{equation}
However, it is an undeniable fact that there might exist some sampling gap close to the security bound, although the sampling strategy improve the information entropy. Some parts of boundary may be inaccurate due to such problem. Thus, a re-sampling mechanism is proposed to find out new critical scenarios ignored in the previous search process. Essentially, this issue can be re-stated as an optimization problem that maximizes the distant to the existing sampling data closest to the stability boundary. Assume that $N_N$ data points $\{(u_{j1},...,u_{j\iota},...u_{j(N_G-1)})\}_{j=1,2,...,N_N}$ have been generated, and $N_{N(\text{new})}$ new points $\{(\widetilde{u}_{\varrho1},...,\widetilde{u}_{\varrho\iota},...\widetilde{u}_{\varrho(N_G-1)})\}_{\varrho=1,2,...,N_{N(\text{new})}}$ are required in the sampling gap area. Thus, the optimization problem is:
\vspace{-3pt}
\begin{equation}
    \label{eq:op0}
    \begin{array}{cl}
        \mathop {\max}\limits_{\widetilde{u}} & \sum\limits_{\varrho=1}^{N_{N(\text{new})}} \left[  \mathop {\min}\limits_{j} \sum\limits_{\iota=1}^{N_{G}-1} (\widetilde{u}_{\varrho\iota}-u_{j\iota})^2 \right] \\

        \text{s.t.} &
        \left
        \{
        \begin{array}{l}
            \varGamma(\widetilde{u},k)=0  \\
            \underline{\widetilde{u}} \le \widetilde{u} \le \overline{\widetilde{u}}\\
        \end{array}
        \right. 	
    \end{array}
    \vspace{-3pt}
\end{equation}
where $j \in [ 1,2,...,N_N ]$, $\iota \in [ 1,2,...,N_{G}-1 ]$, and $\varrho \in [ 1,2,...,N_{N(\text{new})} ]$.

As we can see from the equation shown above, it is a min-max optimization problem which is hard to solve. To simplify, an auxiliary variable $\gamma$ is introduced into this optimization problem, realizing the chordal decomposition of minimum and maximum problem. Thus, the object function of Eq. (\ref{eq:op0}) can be replaced by Eq.(\ref{eq:op1}) shown as follows:
\vspace{-3pt}
\begin{equation}
    \label{eq:op1}
    \begin{array}{cl}
        \mathop {\max}\limits_{\widetilde{u},\gamma} & \sum\limits_{\varrho=1}^{N_{N(\text{new})}} \gamma_\varrho \\
    
        \text{s.t.} &
        \left.
        \begin{array}{l}
            \gamma_\varrho \le \mathop {\min}\limits_{j} \sum\limits_{\iota=1}^{N_{G}-1} (\widetilde{u}_{\varrho\iota}-u_{j\iota})^2 \\
        \end{array}
        \right. 	
    \end{array}
    \vspace{-3pt}
\end{equation}
In fact, it is not necessary to find out the precise global optimal solution, considering that it is only a step for re-sampling. To reduce the required computational time, the duality gap can be preset a larger value. Meanwhile, more initialization points are selected before solving this optimization problem, considering the limitations of nonlinear optimization algorithm.

\vspace{-8pt}
\subsection{Data Generation Termination Criterion}
After focusing on data samples generation strategy, we should evaluate the termination criterion of the generation process. The auxiliary variable $\gamma$ defined in Eq. (\ref{eq:op1}) measures the distance vector between new sampling data and the existing data. Thus, the minimum element of $\gamma$ can be utilized for data generation termination criterion and is defined as follows:
\vspace{-5pt}
\begin{equation}
    \label{eq:data_generation_termination_criterion}
    \mathop {\min}\limits_{\varrho} \gamma_\varrho \leq \gamma_{\text{cri}}
    \vspace{-5pt}
\end{equation}
where $\gamma_{\text{cri}}$ is a preset termination threshold. A reasonable value is less than 1\% of maximum controllable variables in OPs. Lower threshold improves the accuracy but sacrifices efficiency.

\section{Critical scenarios Selection}
\subsection{Preliminary Search Space Selection}
Normally, the proposed method aims at dealing with the changes of possible OPs and contingencies. However, the overall TSB is far too complicated to cover all possible scenarios, and thus we are trying to refresh TBS periodically by tracking current OP in time. In that case, the search space can be reduced significantly, and satisfied the limitation of online computational burden.

\begin{figure}
	\vspace{-12pt}
	\centering
	\includegraphics[width=2.8in]{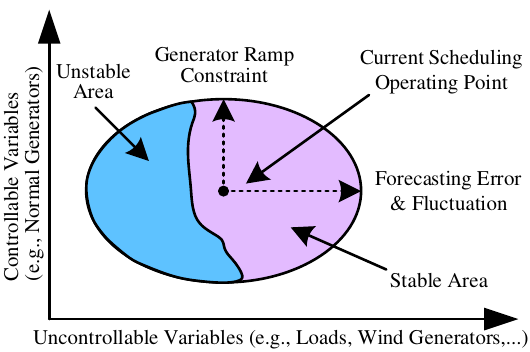}
	\vspace{-8pt}
	\caption{Search space of the current scheduling OP.}
	\label{fig:searcharea}
	\vspace{-15pt}
\end{figure}

As shown in Fig. \ref{fig:searcharea}, the search space can be divided into two main dimensional categories: uncontrollable variables and controllable ones. In terms of the former ones, it mainly includes all kinds of loads, uncontrollable generators (like photovoltaic and wind generators). The exact value can be forecast day (hour or minute) ahead with the forecasting error less than 3\% using state-of-the-art technique. As a result, variables in this part can be sampled randomly within a given small range. In terms of the latter category, it includes most normal generators. Considering the ramping constraint of each generator, the controllable variables are also be limited in a small range (although it is significantly larger than uncontrollable ones). Noted that the search area has a direct relationship with model periodic refreshing frequency. Overall, this process reduces the search space according to the latest scheduling OP.

\vspace{-8pt}
\subsection{Critical Operating Point Selection}
After determining the preliminary search space, the data relevance of all possible OPs are required to be analyzed. First of all, the contingency is assumed to be the same. Under such circumstances, we define a matrix $\bm{\Psi}^k$ with respect to the impact on different OPs $u$ under the same given contingency $k$ using stability index, that is, $\bm{\Psi}^k = \triangledown_\Upsilon \bm{\Phi}^k(u)$ equals to:
\vspace{-3pt}
\begin{equation}
    \label{eq:cluster_matrix1}
    \footnotesize
    [
    \triangledown_\Upsilon \bm{\Phi}^k(u_1), 
    \triangledown_\Upsilon \bm{\Phi}^k(u_2),
    \cdots,
    \triangledown_\Upsilon \bm{\Phi}^k(u_\varsigma),
    \cdots,
    \triangledown_\Upsilon \bm{\Phi}^k(u_{N_u})
    ]^T 
    \vspace{-3pt}
\end{equation}
where $\Upsilon$ is the controllable operation variable (e.g., controllable generators), $\epsilon$ and $N_\Upsilon$ are respectively the index and the dimension of variable $\Upsilon$. $u$ is the OPs, $\varsigma$ and $N_u$ are the index and the dimension of variable $u$. $k$ denotes the index of the given contingency. $\bm{\Psi}^k$ is a $N_u \times N_\Upsilon$ matrix, and the $\varsigma$-th element $\triangledown_\Upsilon \bm{\Phi}^k(u_\varsigma)$ has $N_\Upsilon$ column, and can be described as:
\vspace{-3pt}
    \begin{equation}
    \label{eq:cluster_matrix2}
    \footnotesize
    [
    \triangledown_{\Upsilon_1} \bm{\Phi}^{k}(u_\varsigma), \triangledown_{\Upsilon_2} \bm{\Phi}^{k}(u_\varsigma),
    \cdots,
    \triangledown_{\Upsilon_{\epsilon}} \bm{\Phi}^{k}(u_\varsigma),
    \cdots,
    \triangledown_{\Upsilon_{N_\Upsilon}} \bm{\Phi}^{k}(u_\varsigma)
    ] 
\vspace{-3pt}
\end{equation}

In order to measure the similarity of each OP, Spearman Correlation (SC) Algorithm is employed to classify all possible OPs into several clusters, based on the response of generator rotor angles for a given OP. To calculate the SC value, the matrix defined in Eq. (\ref{eq:cluster_matrix1}) have to be converted into rank vector. For example, the $r$-th row of $\bm{\Psi}^k$ indicates the derivative information of $r$-th OP, and the controllable variable corresponding to the $\rho$-th lowest value is assigned rank $\rho$. Based on this definition, we can calculate the value of SC between the $r$-th and $s$-th OP by the following equation.
\vspace{-3pt}
\begin{equation}
    \footnotesize
	\label{eq:sc}
	SC(r,s)=\frac{\sum^{N_\Upsilon}_{\rho=1}(R_{u_r}(\rho)-\overline{r_{u_r}})(R_{u_s}(\rho)-\overline{r_{u_s}})}{\sqrt{\sum^{N_\Upsilon}_{\rho=1}(R_{u_r}(\rho)-\overline{r_{u_r}})^2 \cdot \sum^{N_\Upsilon}_{\rho=1}(R_{u_s}(\rho)-\overline{r_{u_s}})^2}}
	\vspace{-3pt}
\end{equation}
where $\overline{r_{u_r}}$ and $\overline{r_{u_s}}$ are the average value of rank vector with respect to the $r$-th and $s$-th OP, and are defined as follows.
\vspace{-3pt}
\begin{equation}
	\label{eq:sc_sub}
	\overline{r_{u_r}}=\frac{1}{N_\Upsilon}\sum^{N_\Upsilon}_{\rho=1}(R_{u_r}(\rho)),~~~
	\overline{r_{u_s}}=\frac{1}{N_\Upsilon}\sum^{N_\Upsilon}_{\rho=1}(R_{u_s}(\rho))
	\vspace{-3pt}
\end{equation}

After we got the SC value, spectral clustering technique is employed to classify the contingencies into several clusters. The most severe OP in each cluster can be regarded as the representative of others in the corresponding cluster. 
\begin{figure}[t]
	\vspace{-12pt}
	\centering
	\includegraphics[width=3.5in]{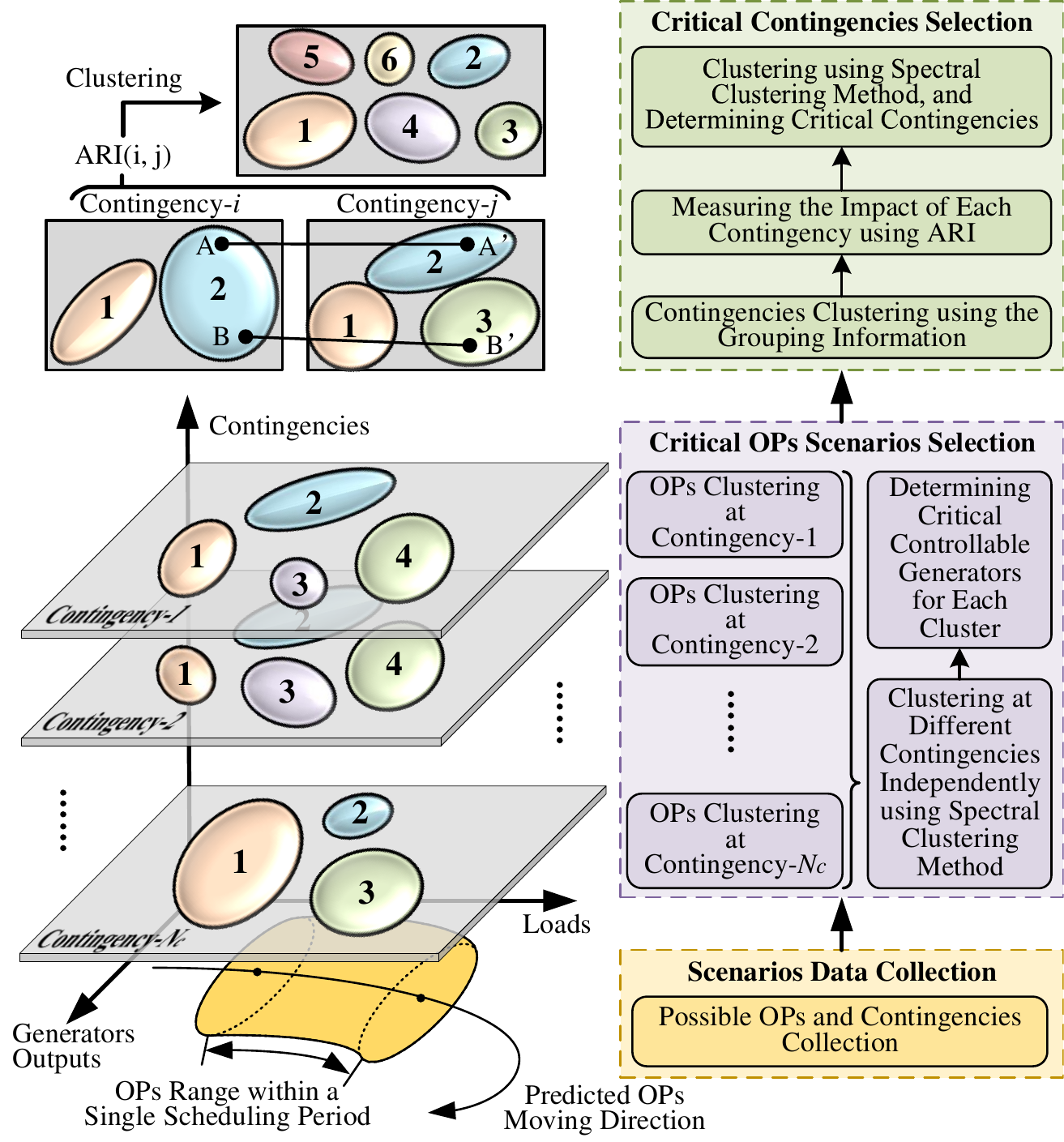}
	\vspace{-15pt}
	\caption{Flowchart of critical scenarios selection.}
	\label{fig:cluster}
	\vspace{-12pt}
\end{figure}

\vspace{-8pt}
\subsection{Critical Contingency Selection}
Considering that the power system may undergo all kinds of contingencies, all possible disturbances have to be taken into account. As discussed in the previous subsection, grouping information has been obtained under the same contingency. In this subsection, therefore, we are trying to identify the representative contingencies under various OPs. If the grouping information is similar for two contingencies, only one needs to be analyzed in the next stage. Based on this idea, we employed adjusted rand index (ARI) to represent the relationship between contingencies and then cluster them into several sets. Given an OPs set with $N_u$ elements, and two contingency clusters of these elements, namely $p=\{p_1,...,p_f,...,p_{N_p}\}$ and $q=\{q_1,...,q_g,...,q_{N_q}\}$. The overlap between these two contingency scenarios $p$ and $q$ are summarized in Eq.(\ref{eq:ari_matrix}).
\vspace{-3pt}
\begin{equation}
	\label{eq:ari_matrix}
	\scriptsize
	\begin{array}{c|cccccc|c}
		p/q&q_1&q_2&\cdots&q_g&\cdots&q_{N_p}&\sum_g\\
		\hline
		p_1&n_{11}&n_{12}&\cdots&n_{1g}&\cdots&n_{1N_p}&a_1\\
		p_2&n_{21}&n_{22}&\cdots&n_{2g}&\cdots&n_{2N_p}&a_2\\
		\vdots&\vdots&\vdots&\ddots&\vdots&\ddots&\vdots&\vdots\\
		p_f&n_{f1}&n_{f2}&\cdots&n_{fg}&\cdots&n_{fN_p}&a_f\\
		\vdots&\vdots&\vdots&\ddots&\vdots&\ddots&\vdots&\vdots\\
		p_{N_p}&n_{N_p1}&n_{N_p2}&\cdots&n_{N_pg}&\cdots&n_{N_pN_p}&a_{N_p}\\
		\hline
		\sum_f&b_1&b_2&\cdots&b_g&\cdots&b_{N_p}&n\\
	\end{array}
	\vspace{-3pt}
\end{equation}
where, each entry $n_{fg}$ (the $f$-th row, $g$-th column) denotes the number of OPs in common between $p_f$ and $q_g$. Thus, the ARI between contingency $f$ and $g$ can be calculated as follows:
\vspace{-3pt}
\begin{equation}
	\label{eq:ari}
	ARI(f,g)=\frac{\sum_{fg}C_{n_{fg}}^2-\frac{\sum_{f}C_{a_f}^2 \cdot \sum_{g}C_{b_g}^2}{C_n^2}}{\frac{\sum_{f}C_{a_f}^2+\sum_{g}C_{b_g}^2}{2}-\frac{\sum_{f}C_{a_f}^2+\sum_{g}C_{b_g}^2}{C_n^2}}
	\vspace{-3pt}
\end{equation}
where $C$ denotes the combinatorial operator.

Similar contingency clustering process is carried out as critical OPs selection. So far, several critical contingencies are selected to reduce the computational burden.

\vspace{-8pt}
\subsection{OP Matching and Critical Generator Selection}
Although the aforementioned approaches have declined the number of OPs significantly, the real OPs vary from the ones in the day-ahead scheduling period, and the OPs matching mechanism is required. It is noted that the most critical generators (MCGs), in other words, the controllable generators that have the most significant impact on system transient stability, are almost the same in each OPs cluster. Therefore, MCGs can be determined once the new OP is matched into one certain cluster. We introduce the multivariate Gaussian model to do so. Suppose $N_u$ OPs under each contingency, the parameter $u_\mu$ and $u_\Sigma$ of a given cluster can be calculated:
\vspace{-3pt}
\begin{equation}
    \label{eq:gau_mu}
    u_\mu=\frac{1}{N_u}\sum_{\varsigma=1}^{N_u}u_\varsigma
    ,~~~
    u_\Sigma=\frac{1}{N_u}\sum_{\varsigma=1}^{N_u}(u_\varsigma-u_\mu)(u_\varsigma+u_\mu)^T
    \vspace{-3pt}
\end{equation}
Then, the possibility of a new OP $u_{\text{new}}$ in cluster $c$ is:
\vspace{-3pt}
\begin{equation}
    \label{eq:gau_p}
    p_c(u_{\text{new}})=\frac{\exp(-\frac{1}{2}(u_{\text{new}}-u_\mu)^Tu_\Sigma^T(u_{\text{new}}-u_\mu))}{\sqrt{(2\pi)^{N_\Upsilon}|u_\Sigma|}}
    \vspace{-3pt}
\end{equation}
Lastly, the new OP is regarded to be in the cluster with largest possibility calculated in Eq.(\ref{eq:gau_p}), and MCGs can be chosen accordingly.

\section{Online Security Monitoring Framework}
Noted that transient stability boundary generation task is an NP hard problem, as the whole possible space of OPs are required to be discretized with small interval. Meanwhile, problem scale increases exponentially with the growth of power system interconnection, considering all possible discretized OPs together with all kinds of contingencies. It is definitely impossible to trace the whole TSB by classical brute force method, especially for power systems with more than 100 buses, before the advent of commercial quantum computers. Although gradient based sampling method is proposed in the previous section to improve the efficiency, search space still keeps the same. As a result, it is still hard to cover all critical scenarios, in order to generate an accurate boundary.
\begin{figure}
	\vspace{-12pt}
	\centering
	\includegraphics[width=3.5in]{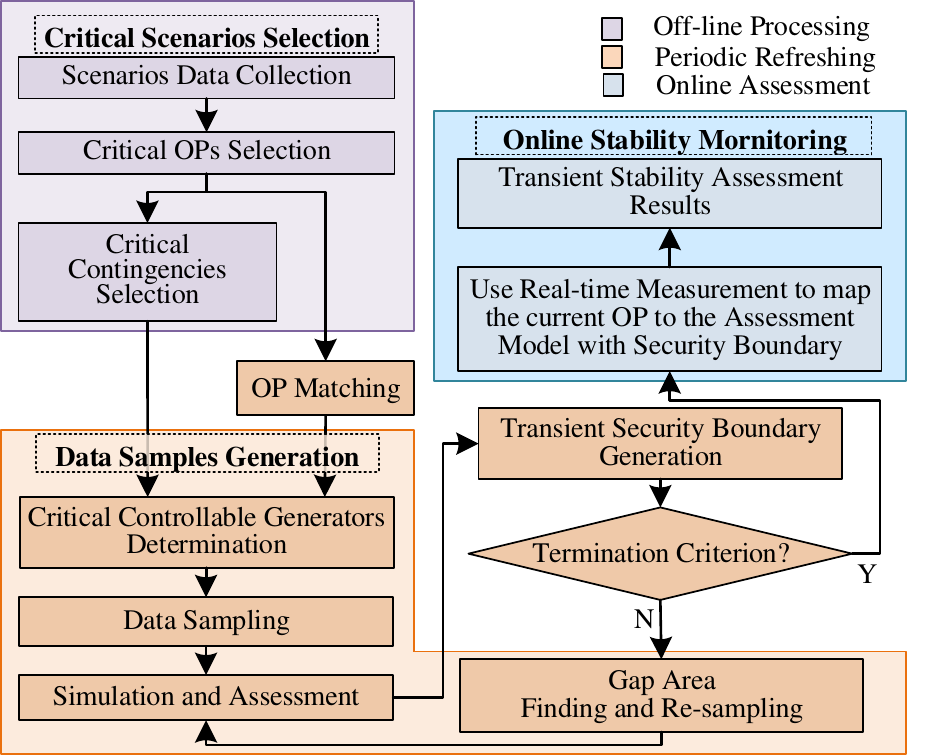}
	\vspace{-12pt}
	\caption{Flowchart of the proposed online security monitoring framework.}
	\label{fig:framework}
	\vspace{-15pt}
\end{figure}

To relieve such difficulties, we introduce a new framework that only several controllable generators which affect stability most are included in the search space of a single scheduling period. Meanwhile, the search space can be further reduced by tracking the current or predicted OPs with relatively small refreshing time interval, in order to reflect the influence of other variables on TSB in time. Considering controllable generators which affect stability most vary from time to time according to different OPs and contingencies, scenarios clustering and matching technique is introduced in this step.

Based on such idea, we propose a new algorithm flow which is implemented as Fig. \ref{fig:framework}. It is observed that, in this figure, the whole process can be divided into three parts based on time scales: off-line precessing, periodic refreshing and on-line assessment. For the first stage, large amount of data with various scenarios are collected, and the most critical OPs and contingencies are selected, in order to make a preparation for most critical controllable generators selection process in the next stage. It is noted that this stage is executed off-line, and aims at reducing the search space. In terms of the second stage, the current OPs are matched with existing clustering result using multivariate Gaussian model. Meanwhile, critical controllable generators can be determined based on the matched scenarios, which significantly affect transient stability. After determining the critical controllable generators and scenarios, data are sampled and generated using the proposed gradient based method, to generate or refresh the accurate boundary. Meanwhile, data points are also re-sampled and generated until reaching the termination criterion. So far, an accurate TSB based on the current scheduling period has been obtained, and can be utilized in the online assessment stage. It is noted that the boundary is updated continuously by tracking the current OP, ensuring the accuracy of online assessment.

\section{Case Studies}
Two typical systems with different scales are investigated in this section: IEEE 9-bus test system and NESTA 162-bus system.  All cases are tested on a computer with Intel Core i7-4790 3.6GHz CPU, 16GB RAM unless otherwise specified.

\vspace{-8pt}
\subsection{Visualize the Generation Process: IEEE 9-bus Test System}
\noindent\textit{1) Test System and Configurations}

In the first case study, the proposed algorithm is applied on a small system: IEEE 9-bus test system. The steady-state and transient parameters are referred in MATPOWER \cite{matpower} and PSAT manual \cite{psat}, respectively.

Considering that there are only 3 generators in this grid, it is still possible to visualize the generative process of TSB without dimension reduction and to verify all possible OPs that violate the security constraints of the given system. More specifically, it is highlighted in this subsection that the proposed algorithm is able to generate more data samples close to the TSB, in order to improve efficiency without compromising on accuracy of the boundary. Therefore, we measure not only efficiency but accuracy improvement visually compared with several existing methods.

In this study, all loads are assumed to change randomly and independently within $\pm$10$\%$ of its reference level. The contingency preset is initialed by a three-phase to ground fault at bus 5, and cleared after 0.2 seconds by tripping a line between bus 5 and 7. The stability performance is assessed using TDS and determined by maximum rotor angle difference. For comparison, accurate TSB is generated by brute force, that is, performing TDS at all possible OPs with discretized interval of 1MW.

\noindent\textit{2) Visualized Result of the Generation Process}

\begin{figure}
	\vspace{-12pt}
	\centering
	\includegraphics[width=3.5in]{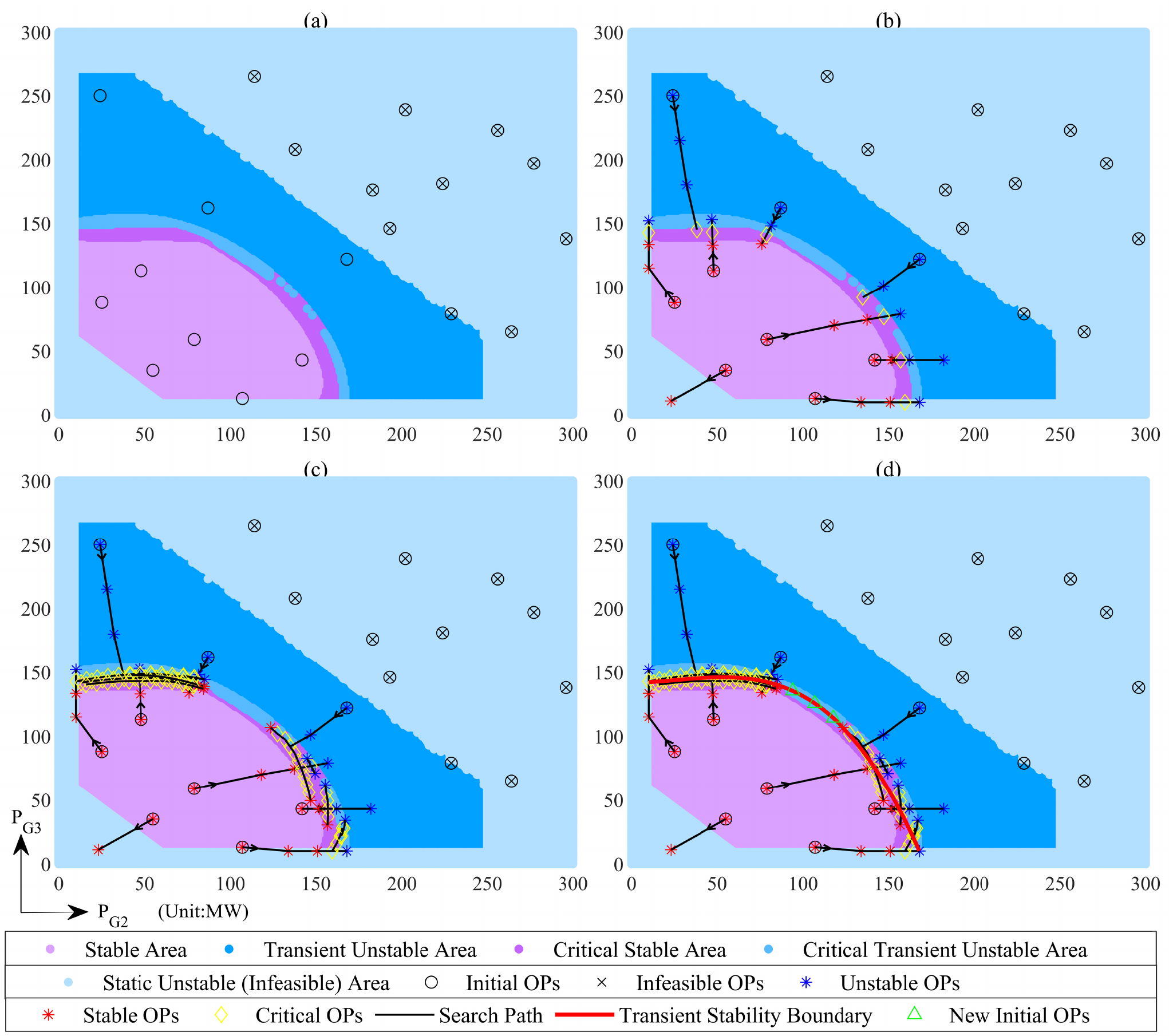}
	\vspace{-15pt}
	\caption{Result of search path and sampled data points based on critical data sampling framework.}
	\label{fig:case9_all}
	\vspace{-12pt}
\end{figure}
As shown in Fig. \ref{fig:case9_all}(a), 20 initial OPs are randomly selected using Latin Hypercube Sampling method within the given output range of generators. Among them, 11 OPs with the '$\otimes$' mark are found in the infeasible area by static security check after solving power flow. The remaining 9 OPs are regarded as the initial seeds to generate the rest samples. 
Fig. \ref{fig:case9_all}(b) shows the process of getting close to the boundary using specially designed critical data sampling strategy introduced in Section II. It can be easily found that the step size varies according to the distance to the boundary. If the sample is getting closer to the boundary, the next step size becomes smaller. This ensures that the sampling near the boundary is sufficient, while traversing the rest less-informative part with fewer samples. Occasionally, some search processes cross the boundary due to the highly nonlinear nature of TSB, binary search is then employed to recover from such cases.

Additionally, the new transient samples can also be searched perpendicular to the gradient direction based on the existing samples near the boundary, as shown in Fig. \ref{fig:case9_all}(c). However, it can be observed that there exists the sampling gap on part of TSB. Therefore, more data samples in that area are required to generate a more accurate boundary. A rough boundary is generated as the red line and new sampled points are gotten in the gap area as green triangle markers in Fig. \ref{fig:case9_all}(d). 

So far, new sampled points can be regarded as the new initial seeds and repeat the above procedures, so as to generate a more accurate boundary with the increase of data samples.

\noindent\textit{3) Gradient Information for Possible OPs}

\begin{figure}
	\vspace{-4pt}
	\centering
	\includegraphics[width=3.5in]{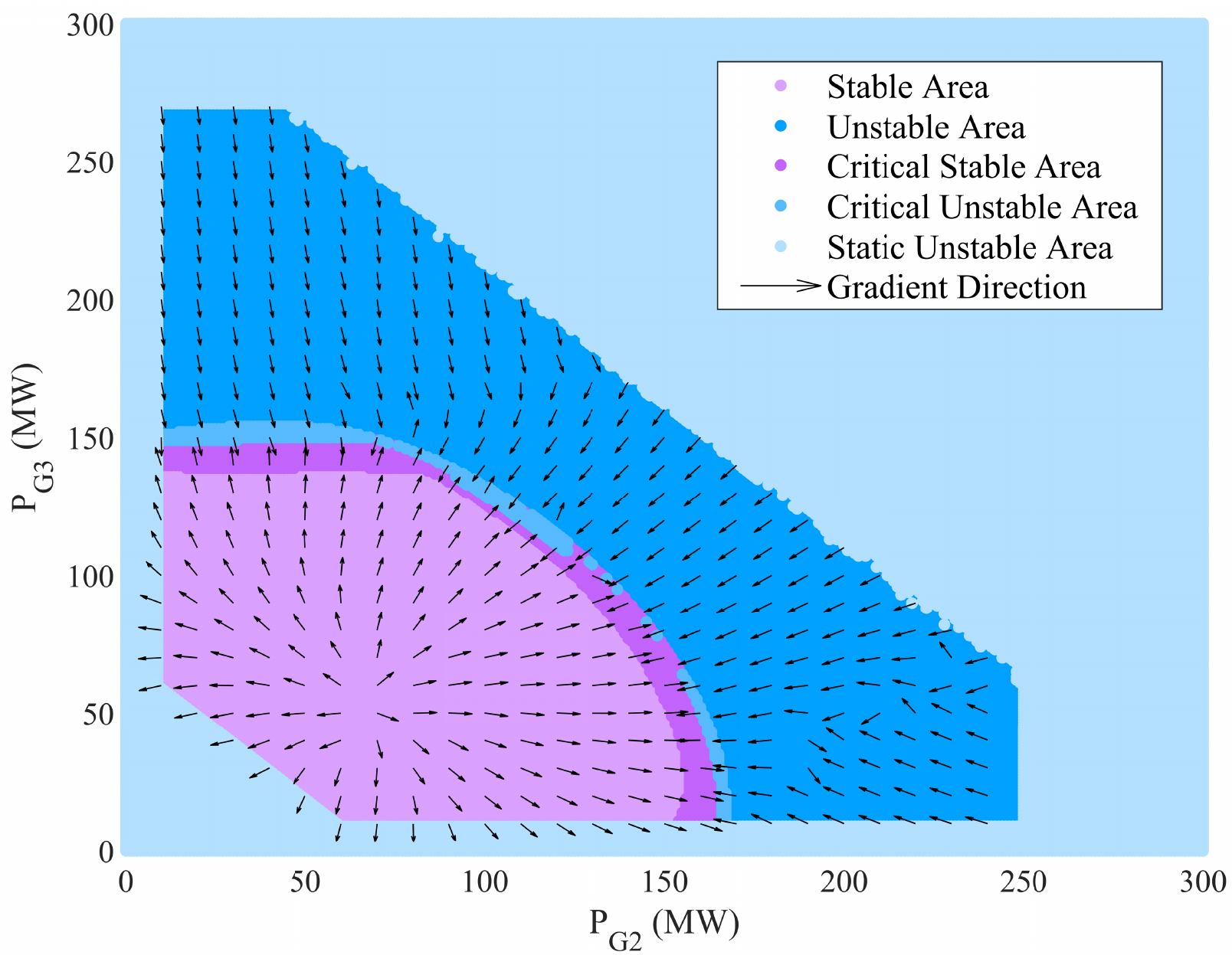}
	\vspace{-15pt}
	\caption{Gradient direction plot for all OPs in static stable area.}
	\label{fig:case9_gradient}
	\vspace{-12pt}
\end{figure}

Fig. \ref{fig:case9_gradient} shows gradient direction information for all possible OPs in static stable area, with discretization interval of 5MW. It is observed that almost all gradient direction arrows point to transient stability boundaries using the proposed transient index and algorithm. Therefore, it proves effective to generate more data samples close to the stability boundary using this specially designed index and method.

\noindent\textit{4) Numerical Results of the Proposed Method}

Throughout the whole process of the data and boundary generation, the proposed method shows significant effect on efficiency without compromising accuracy. We compare the proposed method with four existing ones, summarized in Table \ref{tab:case9_compare} in details. For comparison more conveniently, we set a minimum accuracy requirement for all methods. Enabled by the proposed method, the data size required to reach 99.9\% accuracy declines by nearly 80\% and the time required reduces by 32\%, compared with the most state-of-the-art method (importance sampling based method). It is highlighted that the proposed method shows better performance.

\begin{figure}
	\centering
	\includegraphics[width=3.5in]{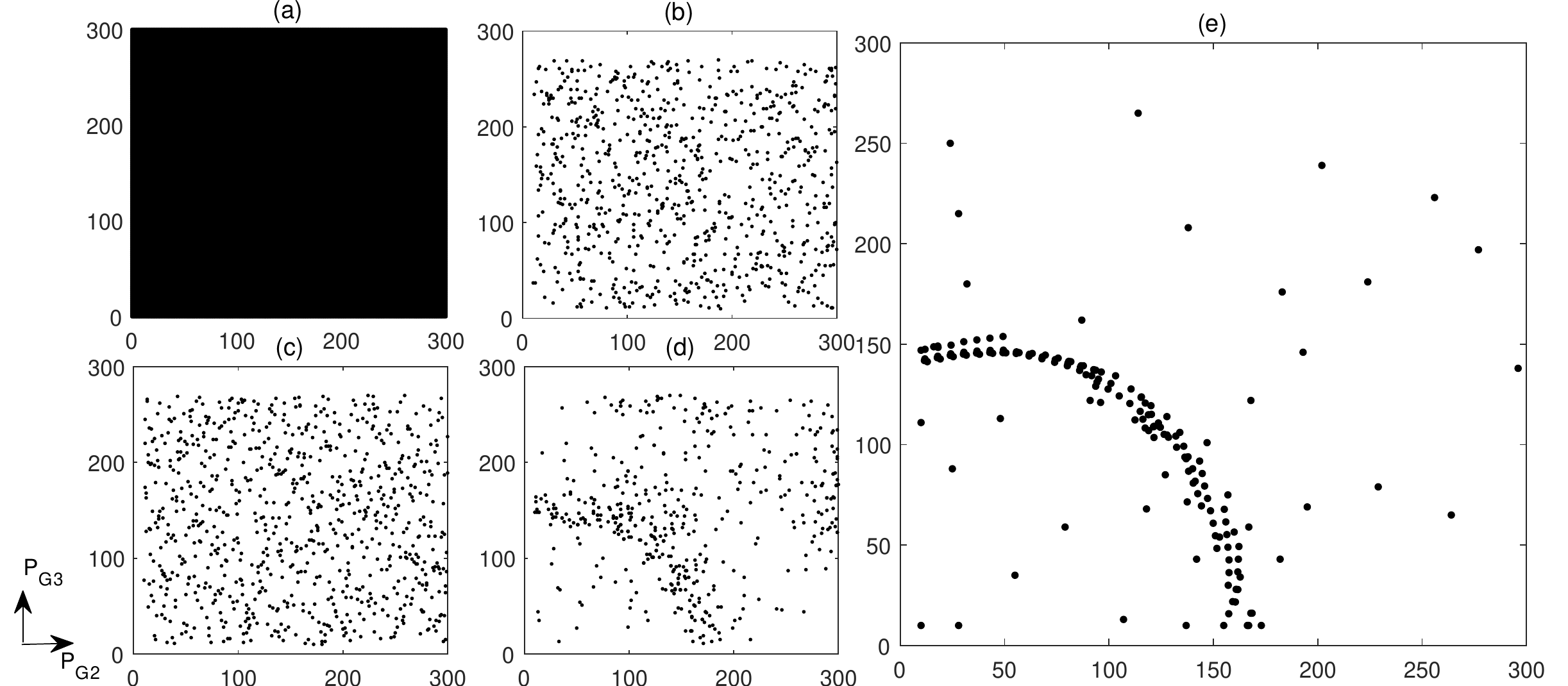}
	\begin{tablenotes}
		\scriptsize
		\item *(a) brute force method, (b) random sampling method, (c) latin hypercube sampling method, (d) importance sampling method, (e) proposed method.
	\end{tablenotes}
	\vspace{-4pt}
	\caption{Sampling OPs scatter diagram using different methods.}
	\label{fig:case9_ex2}
	\vspace{-15pt}
\end{figure}

Meanwhile, the scatter plot of the data samples using different methods is illustrated in Fig. \ref{fig:case9_ex2}. We can easily distinguish the stability boundary through the last two plots (importance sampling based and the proposed methods) rather than the first three. Among them, the data samples generated by the proposed method are much closer to the boundary, and thus shows the superior performance.

\begin{table}
	\caption{Indices Comparison between the Proposed and Existing Methods}
	\label{tab:case9_compare}
	\vspace{-6pt}
	\centering
	\scriptsize
	\begin{tabular}{crrrrrrr}
		\toprule
		\multirow{1}{*}[-0pt]{\begin{tabular}{@{}c@{}}Method\end{tabular}}
		&\multirow{1}{*}[-0pt]{\begin{tabular}{@{}c@{}}Data Size\end{tabular}}
		&\multirow{1}{*}[-0pt]{\begin{tabular}{@{}c@{}}Accuracy(\%)\end{tabular}}
		&\multirow{1}{*}[-0pt]{\begin{tabular}{@{}c@{}}Time(s)\end{tabular}}
		
		\\
		\midrule
		Brute Force               &  $90,601$  &  $100$   & $2,084.00$   \\
		Ramdom Sampling           &  $800$  &  $99.90$   & $19.33$   \\
		Latin Hypercube Sampling  &  $800$  &  $99.91$  & $19.98$   \\
		Importance Sampling        &  $500$  &  $99.91$  & $14.09$   \\
	    Proposed Method           &  $117$  &  $100$  & $9.59$   \\
		
		\bottomrule
	\end{tabular}	
	\begin{tablenotes}
		\item *All possible OPs (90,601 OCs) in this case are taken into consideration to evaluate the accuracy performance of each approach.
		\item **The OPs in critical area can be seen as either stable or unstable points.
		\item ***In terms of existing methods, more samples result in higher accuracy and lower efficiency. For comparison, the minimum accuracy requirement is set to 99.9\%.
	\end{tablenotes}
	\vspace{-10pt}
\end{table}

\vspace{-6pt}

\vspace{-8pt}
\subsection{A Higher Dimension System: NESTA 162-bus System}

\noindent\textit{1) Test System}

In the second case study, the proposed method is applied to a larger and more complex power system named NESTA 162-bus system. Considering the large number of controllable generators and possible "N-1" contingencies in this grid, computational burden increases geometrically in order to generate TSB. Therefore, we focus on the most critical scenarios selection under different circumstances, to ensure that the online computational burden is under control.

\noindent\textit{2) Most Critical OPs and Contingencies Selection}

1,000 OPs are selected randomly according to the load prediction and dispatching plan within the given scheduling period. Meanwhile, 512 contingency, which is initiated by a three-phase-to-ground fault at any line close to bus of one end and cleared after 0.2 seconds by tripping the line, is also selected on all these 1,000 OPs. So far, 512,000 samples are selected based on day-ahead scheduling, to find out the most representative scenarios in the next scheduling period.

\begin{figure}[htbp]
	\vspace{-5pt}
	\subfigure[The analysis of cluster number using eigenvalues.]
	{
		\centering
		\includegraphics[width=3.4in]{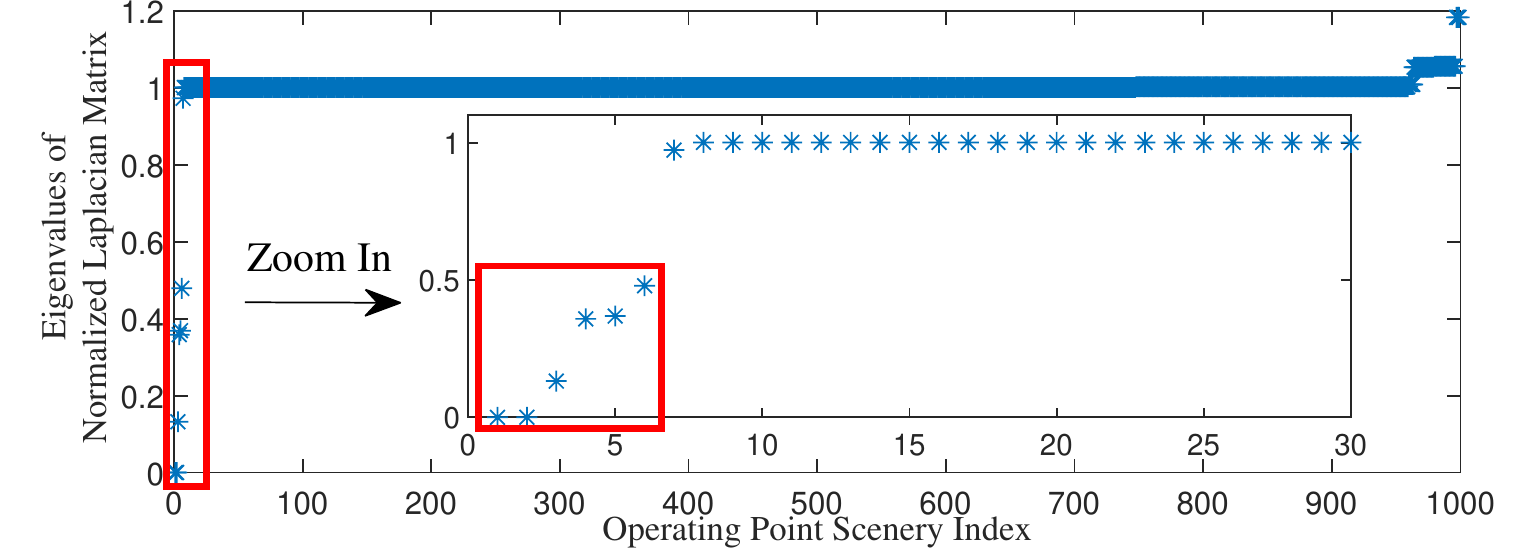}   
		\vspace{-12pt}
	}
	\subfigure[OPs clustering.]
	{
		\centering
		\includegraphics[width=1.65in]{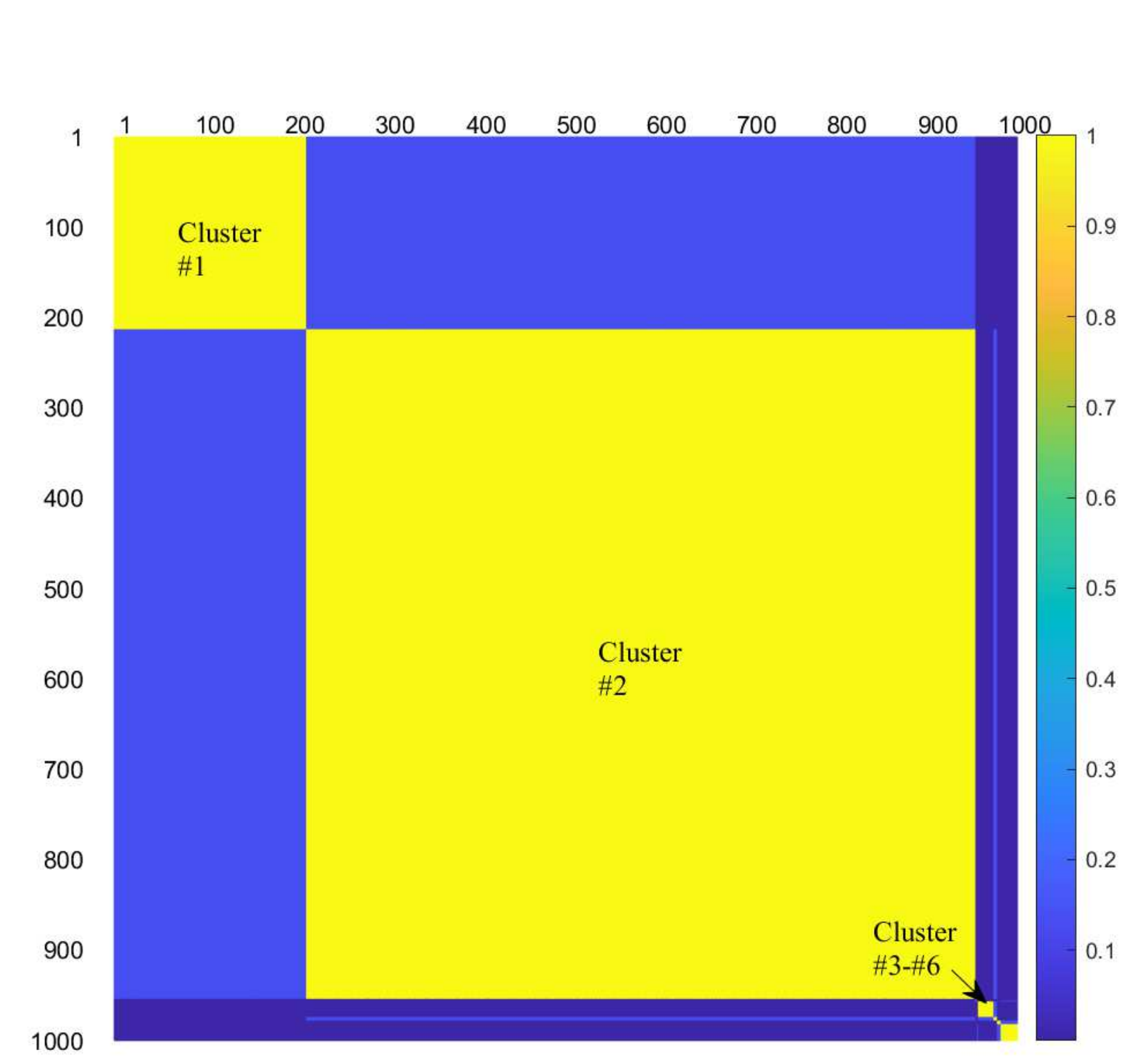}   
	}
	\subfigure[Multi-contingencies clustering.]
	{
		\centering
		\includegraphics[width=1.65in]{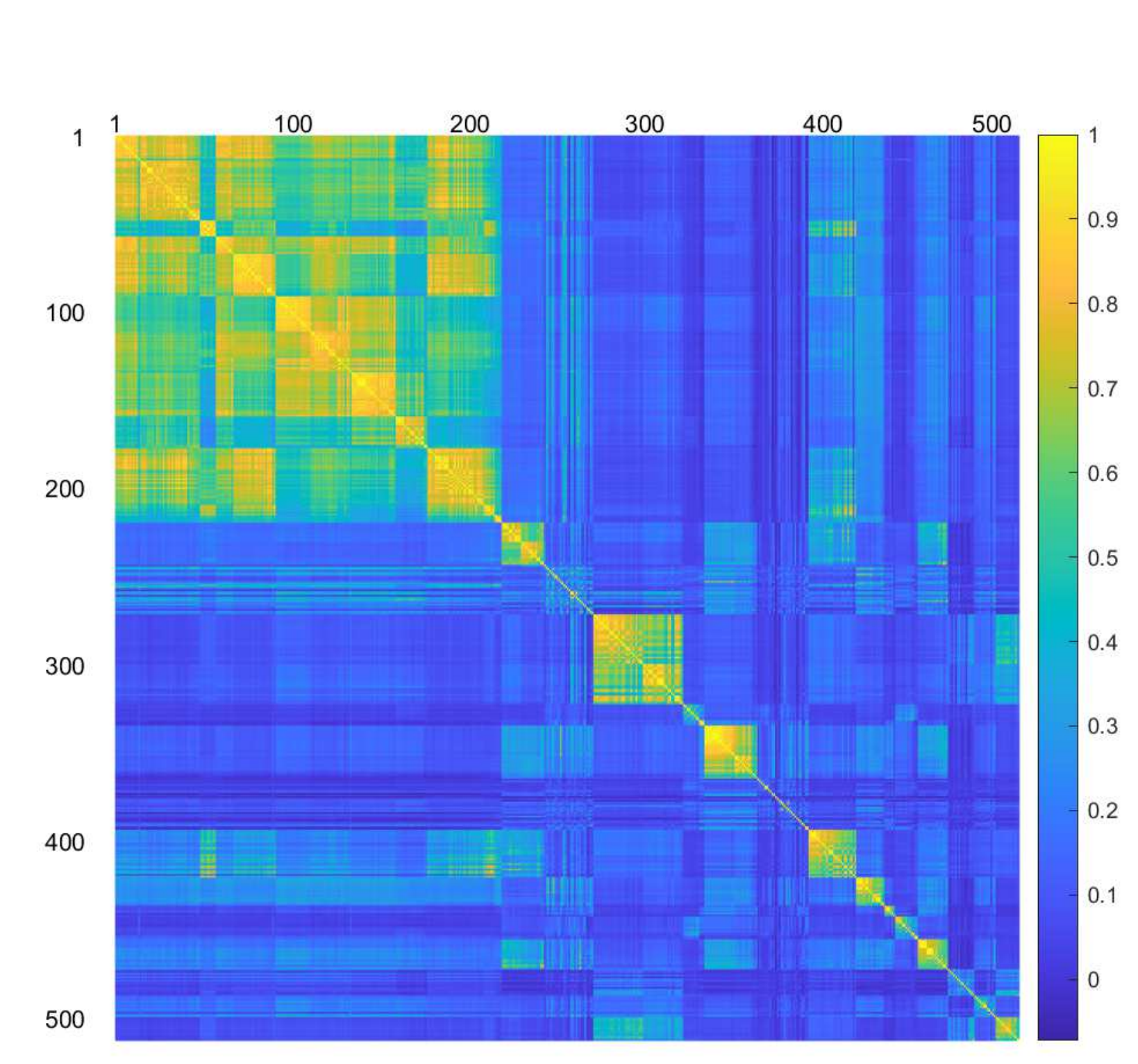}
	}
	\caption{Scenarios (OPs and contingencies) clustering results.} 
	\label{fig:casestudy162_cluster}
	\vspace{-11pt}
\end{figure}

Various OPs under a single contingency are, firstly, to be clustered into several categories. The matrix $\triangledown_\Upsilon \bm{\Phi}\in\Re^{1000\times12}$ as Eq.(\ref{eq:cluster_matrix1}) is constructed, and employed for clustering. Fig. \ref{fig:casestudy162_cluster}(a) illustrates the eigenvalue of normalized Laplacian matrix in spectral clustering of all 1,000 OPs under contingency \#1. It is observed that the first 6 eigenvalues are relatively small, while the others are large. Therefore, cluster number is set to 6. Note that, the number of cluster varies from 2 to 8 for different contingencies. Fig. \ref{fig:casestudy162_cluster}(b) shows the OPs clustering result under contingency \#1. The colored matrix shows the correlation relationship between all different OPs under the same contingency. Darker matrix elements indicate weak correlation between the two OPs, and vice versa. Among them, most OPs belong to the first two clusters, and only small amount of OPs fall into other four clusters. Therefore, only 6 critical OPs are taken into account, since they represents all possible OPs within a given scheduling period. In other words, this reduces the number of OPs dramatically from 1,000 to 6.

After obtaining all grouping and critical OPs information for these 512 contingency scenarios, critical contingencies are also required to be identified. Similar algorithm is applied in this task with result shown in Fig. \ref{fig:casestudy162_cluster}(c). As observed, it can be divided into 25 categories in total. So, only 25 contingencies are required to represent 512 preset contingencies.

Additionally, it is also necessary for us to analyze the efficiency of the clustering process, although it is carried out off-line. The time consuming of OPs clustering is only 2.8-3.0 seconds per each contingency. Considering this step is of natural parallel characteristic, asynchronous parallel algorithm can be employed here if necessary. While the time consuming for contingencies clustering is 24.8 seconds. In a word, it is time-effective to find out the most critical OPs and contingencies, compared to doing TDS for a huge amount of scenarios.

\noindent\textit{3) Test Results of Scenario Matching and Periodic Refreshing}

\begin{figure}
	\vspace{-5pt}
	\centering
	\includegraphics[width=3.4in]{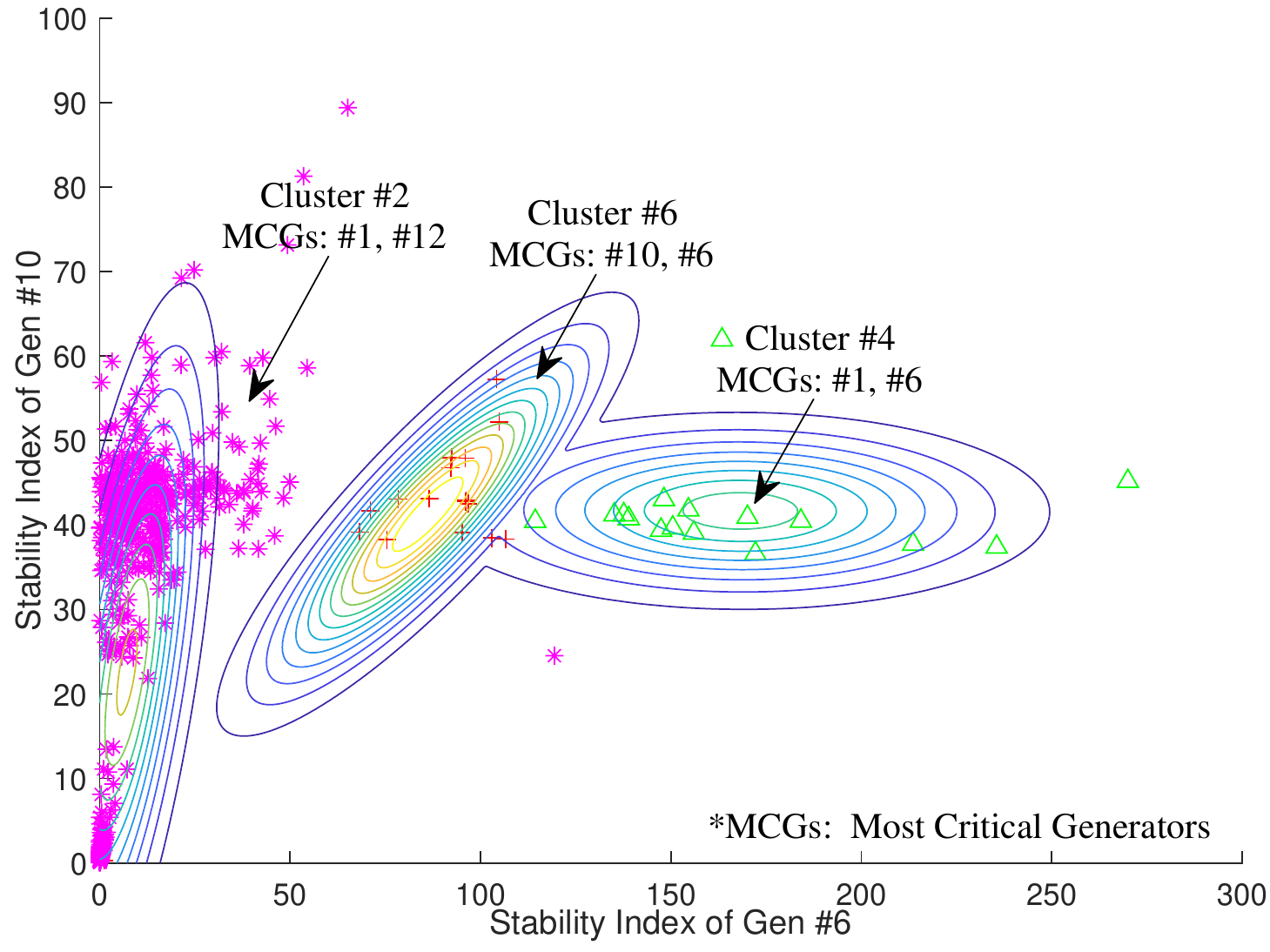}
	\vspace{-10pt}
	\caption{OPs matching and critical generators selection result.}
	\label{fig:case162_gau}
	\vspace{-10pt}
\end{figure}

In on-line operation stage, OP varies from time to time with different circumstances. Although huge amount of data samples are employed off-line, it is still impossible for almost all real OPs to match with the existing samples exactly. Considering that all clusters are difficult to be distinguished on a 2-D plane, we select three of them, shown in Fig. \ref{fig:case162_gau}, as an example to clarify this issue. In this figure, three selected OPs clusters are marked with different colors, together with the probability distribution contour plot. 
As seen in this figure, the dots with magenta asterisk mark belong to cluster \#2, and MCGs in these scenarios are \#1 and \#12. In other words, these two generators are the key for operator to monitor and control in and near the current OP to prevent possible contingencies. Similar results are found for the other clusters except the critical generators. While encountering a new OP in real-world scheduling, multivariate Gaussian distribution probability result is utilized to evaluate the most possible cluster, and to determine MCGs. 1,000 new scenarios are evaluated and 96.1\% of them obtain the same index of MCGs. Although the remaining 3.9\% scenarios are not the most 2 critical generators, they still rank top 3 or 4. Considering the continuously updating in the scheduling period of the proposed method, it has little impact on the security assessment.

\begin{figure}
    \vspace{-6pt}
	\centering
	\includegraphics[width=3.3in]{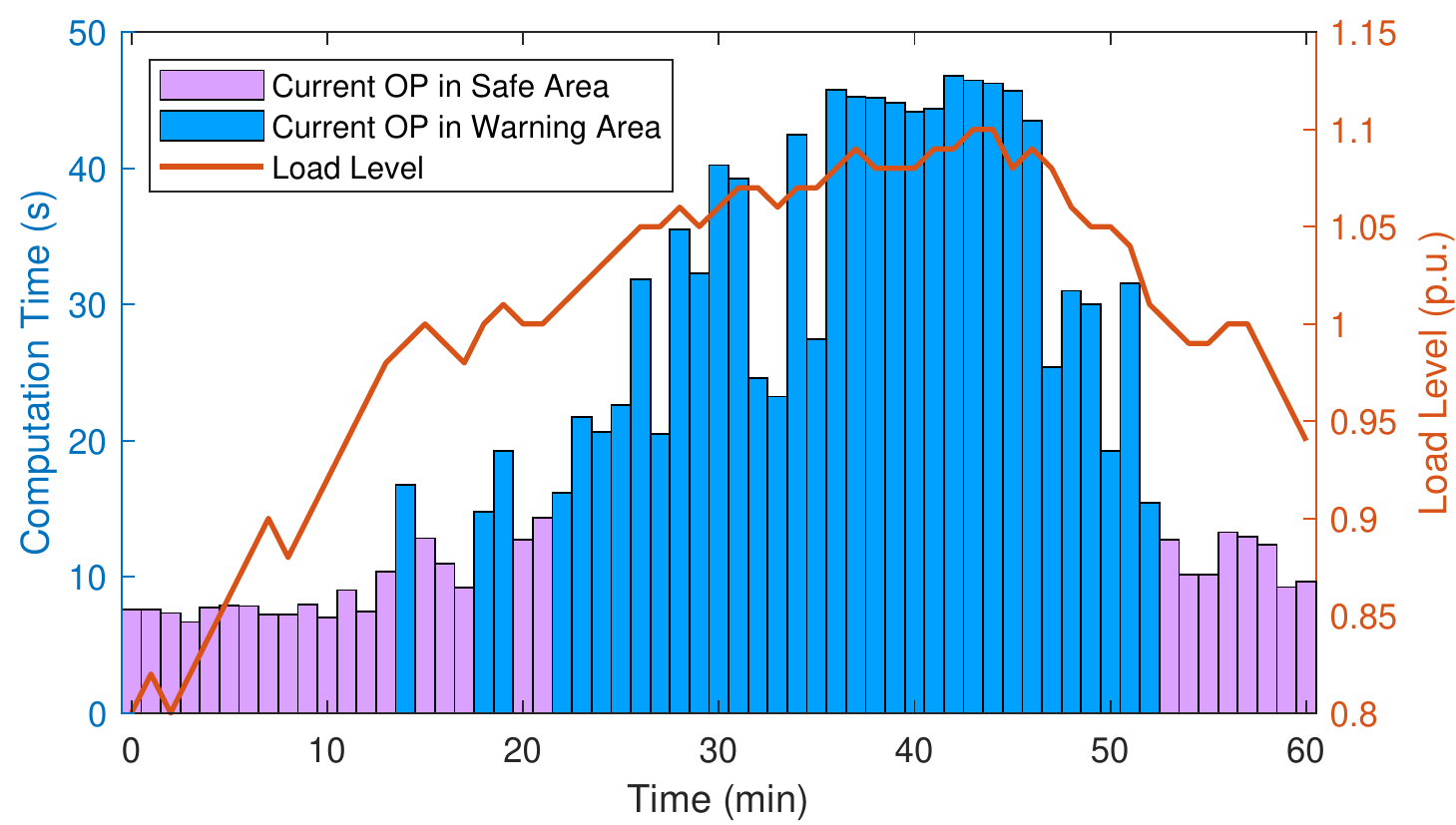}
	\vspace{-5pt}
	\caption{Load level and results of TSB calculation time.}
	\label{fig:casestudy162_loadlevel}
	\vspace{-15pt}
\end{figure}

\begin{figure}
	\centering
	\includegraphics[width=3.5in]{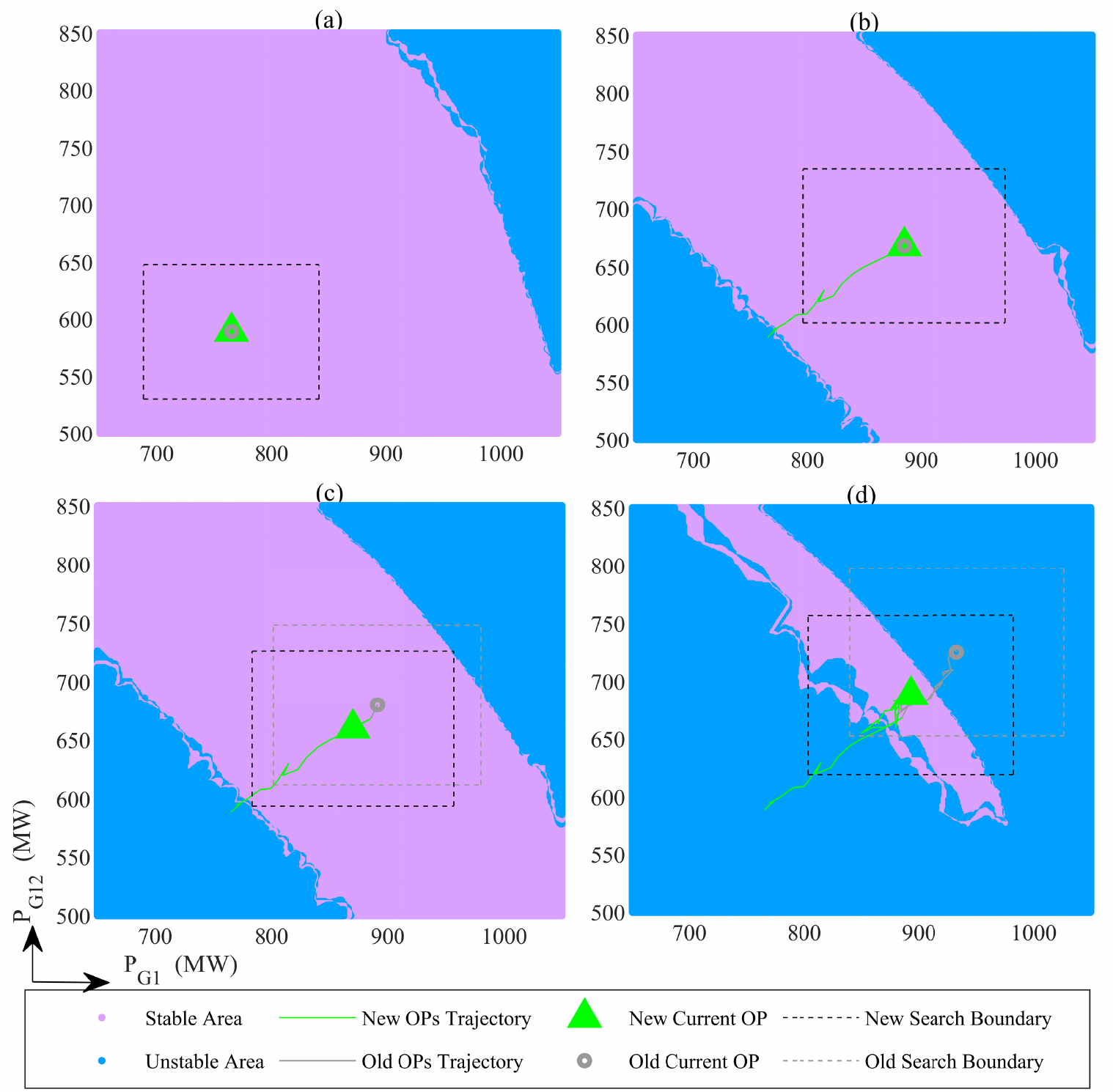}
	\begin{tablenotes}
		\scriptsize
		\item *(a) t=0min, (b) t=14min, (c) t=15min, (d) t=44min.
		\item **Download link of GIF animation is listed in footnote-1 for more time sections. 
	\end{tablenotes}
	\vspace{-5pt}
	\caption{Results of TSB periodically refreshing mechanism.}
	\label{fig:casestudy162_update}
	\vspace{-15pt}
\end{figure}

Besides, in order to evaluate the effect of periodic refreshing mechanism using the proposed algorithm in higher dimensional power system, we generate the dynamically updated TSB in a one-hour scheduling period with its load rate ranging from 0.8 p.u. to 1.1 p.u.(see Fig. \ref{fig:casestudy162_loadlevel}). As shown in Fig. \ref{fig:casestudy162_update}(a), all OPs in the search area (in the area surrounded by dotted line) are stable, when the load level is relatively low. At this stage, it only takes less than 10 seconds to generate the boundary within the search area, and make a conclusion that the state of the current OP is safe. It can be observed that, however, with the increase of load level, some OPs in the search area fall into transient unstable area shown in Fig.\ref{fig:casestudy162_update}(b). As a result, the operator are encouraged to reduce the outputs of generator \#1 and \#12 (see Fig.\ref{fig:casestudy162_update}(c)), while increasing others to some extent. By periodic refreshing the boundary, operators can adjust the current OP continuously, maintaining sufficient stability margin all the time. Even if the load level reaches its maximum as Fig.\ref{fig:casestudy162_update}(d), the OPs still in the stable area. More details can be referred in GIF animation\footnote{Results for more time sections in GIF animation format can be downloaded at
\url{http://genggc.org/files/YanTSB2020.gif}}.

It is noted that the time consuming to generate the TSB is less than 50 seconds (see Fig. \ref{fig:casestudy162_loadlevel} for more details) according to the length of boundary in search area. Thus, TSB can be refreshed every minute. Moreover, parallel technique can be employed in the future to generate data samples, because different search path (see route \circled{1}, \circled{2} and \circled{3} in Fig.\ref{fig:search}) is with the character of naturally parallel. Additionally, it helps in further reducing the refreshing interval to improve the hardware conditions in practical applications.

\section{Conclusions}
This paper has proposed a data-driven transient stability boundary generation framework for online security monitoring. In doing so, a critical data sampling framework and data gap area re-sampling mechanism have been proposed to accelerate the process of generating sufficient informative data samples near and across the boundary. Meanwhile, critical scenario selection strategy is developed to identify the relevance of scenario set and to further reduce the search space of high dimension power systems, enabling the possibility of periodic updating boundary tracking the current OP. The results of case studies illustrated that the proposed method reduces the computational burden of boundary generation process.

In sum, the proposed method offers advantages as follows:

\begin{enumerate}
	\item Improving the efficiency of data generation with most critical scenarios;
	\item Reducing the computational burden of boundary generation and periodic updating by tracking the current OP.
	\item Enhancing the transient stability of the power systems by monitoring TSB and adjusting the current OP. 
\end{enumerate}

\ifCLASSOPTIONcaptionsoff
  \newpage
\fi

\bibliographystyle{IEEEtran}
\bibliography{./bare_jrnl}

\vspace{-15mm}
\begin{IEEEbiography}[{\includegraphics[width=1in]{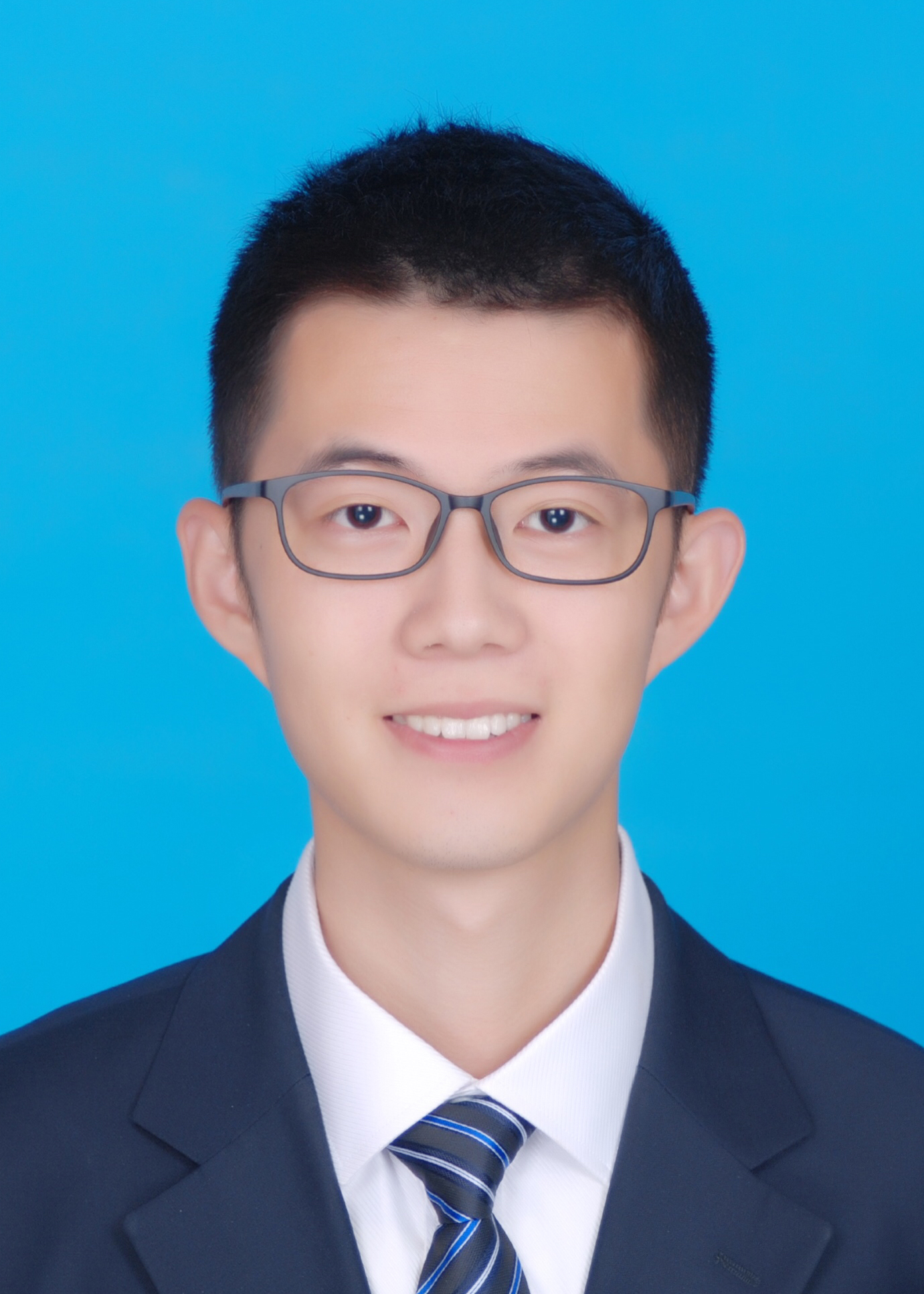}}]{Rong Yan}
	(S'19) received his B.S. degrees in electrical engineering from the School of Electrical Engineering, Wuhan University, Wuhan, China, in 2016. From September 2016 to February 2019, he was a master student at the College of Electrical Engineering, Zhejiang University, Hangzhou, China.
	
	Currently, he is pursuing the Ph.D. degree at the College of Electrical Engineering, Zhejiang University, Hangzhou, China. He is also a visiting student at the Department of Electrical and Computer Engineering, Iowa State University, Ames, United States.
	His research interest is the application of data-driven methods in power system stability analysis.
\end{IEEEbiography}
\vspace{-12mm}
\begin{IEEEbiography}[{\includegraphics[width=1in]{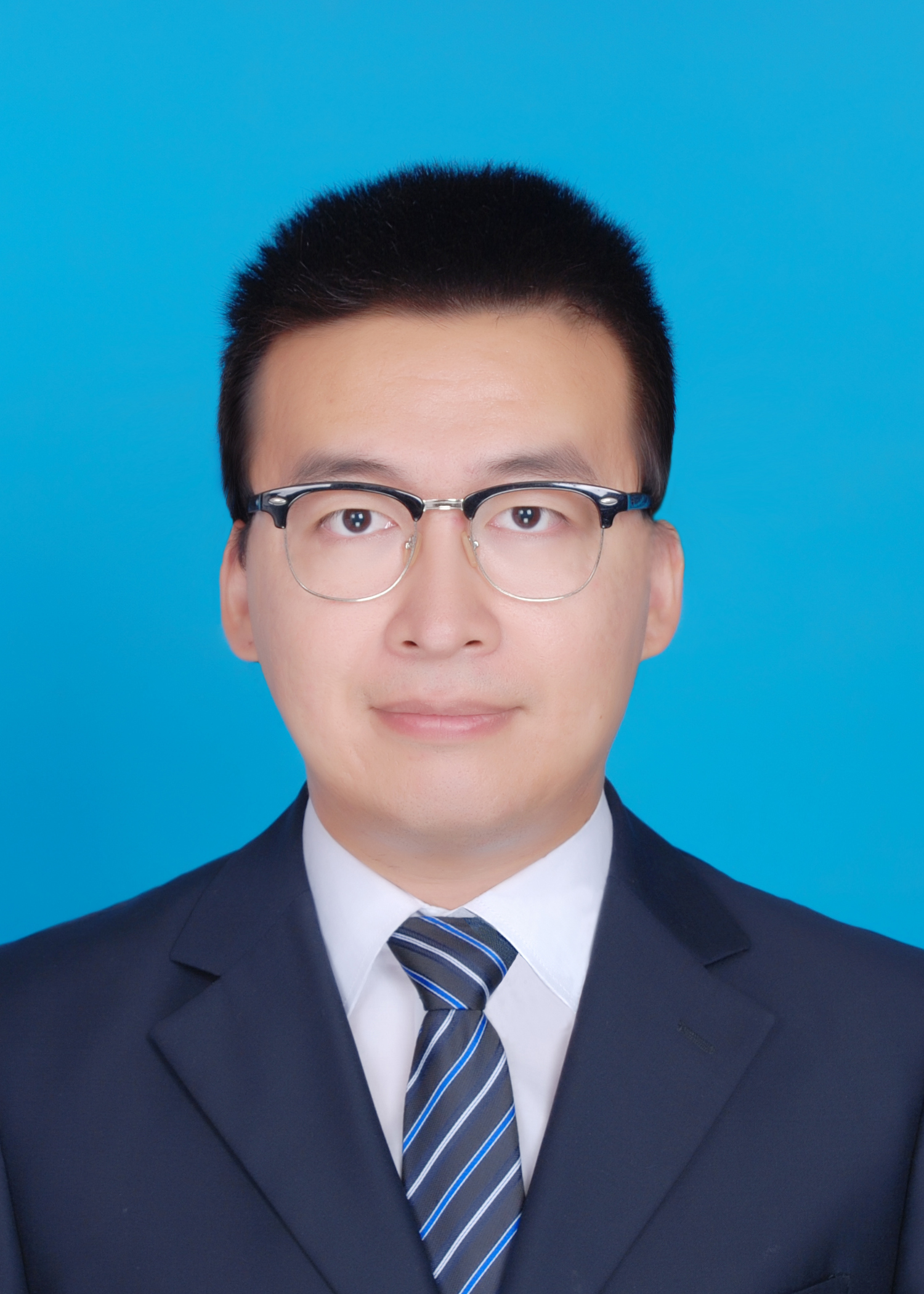}}]{Guangchao Geng}
	(S'10-M'14-SM'19) received his B.S. and Ph.D. degrees in electrical engineering from the College of Electrical Engineering, Zhejiang University, Hangzhou, China, in 2009 and 2014, respectively. From 2012 to 2013, he was a visiting student at the Department of Electrical and Computer Engineering, Iowa State University, Ames, United States. From 2014 to 2017, he is a post-doctoral fellow at the College of Control Science and Engineering, Zhejiang University, Hangzhou, China and the Department of Electrical and Computer Engineering, University of Alberta, Edmonton, AB, Canada.  From 2017 to 2019, he was a research assistant professor at the College of Electrical Engineering, Zhejiang University, Hangzhou, China.
	
	Currently, he is an associate professor at the College of Electrical Engineering, Zhejiang University, Hangzhou, China.
	His research interest includes power system stability and control, the applications of internet of things (IoT) technique and high performance computing (HPC) technique in power systems.
\end{IEEEbiography}
\vspace{-12mm}
\begin{IEEEbiography}[{\includegraphics[width=1in]{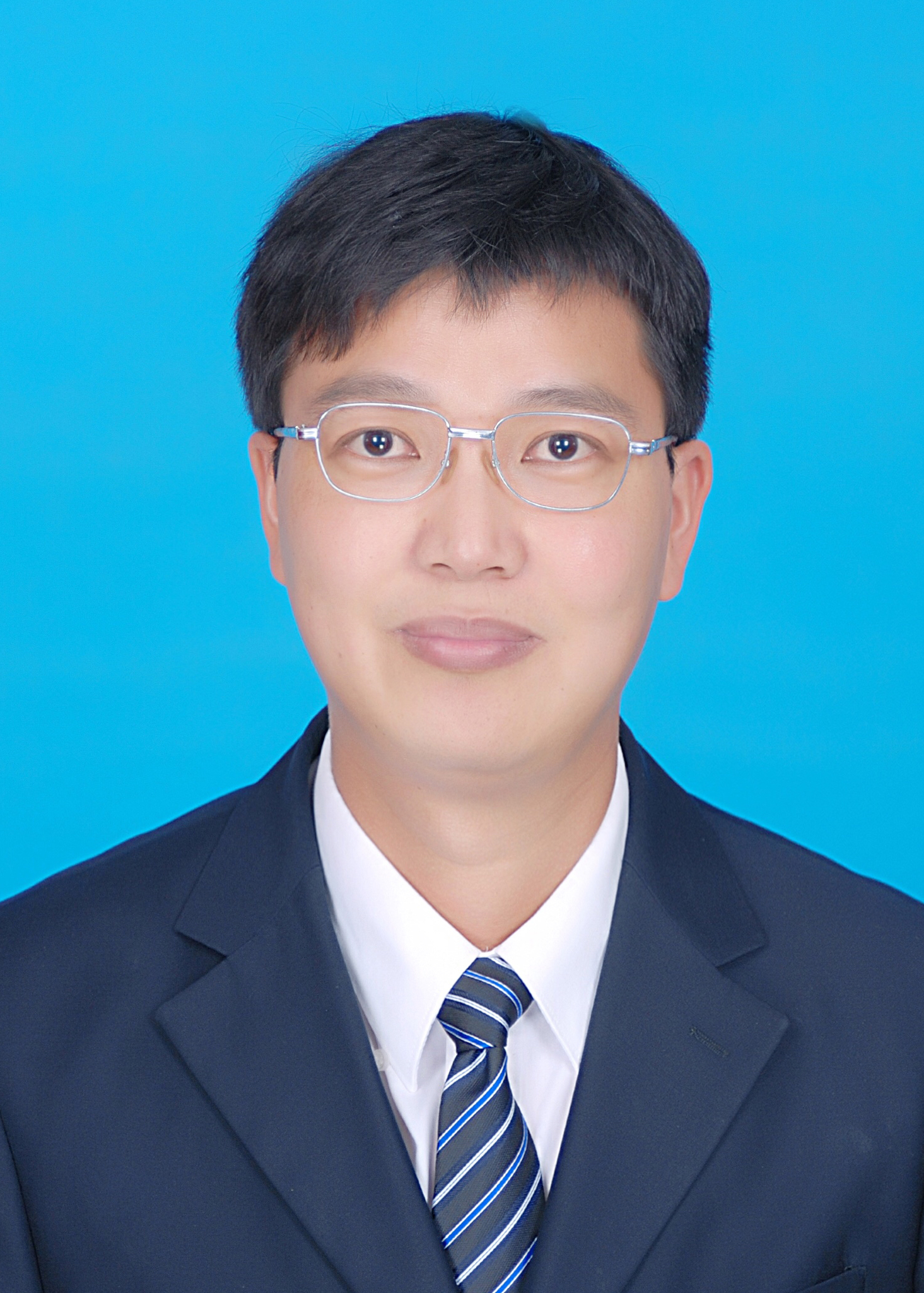}}]{Quanyuan Jiang}
	(M'10-SM'19) received his B.S., M.S., and Ph.D. degrees in electrical engineering from Huazhong University of Science \& Technology, Wuhan, China in 1997, 2000, and 2003, respectively. From 2006 to 2008, he was a visiting associate professor at the School of Electrical and Computer Engineering, Cornell University, Ithaca, United States.
	
	He is currently a professor at the College of Electrical Engineering, academic dean of Graduate School and the vice dean of Polytechnic Institute, Zhejiang University, Hangzhou, China. 
	His research interest includes power system stability and control, applications of energy storage systems and high performance computing technique in power systems.
\end{IEEEbiography} 

\end{document}